\documentclass[
    aps,
    prl,
    reprint,
    preprintnumbers,
    superscriptaddress,
    amsmath,
    amssymb,
    nofootinbib,
    nobibnotes,
    longbibliography
]{revtex4-2}

\usepackage{graphicx}   
\usepackage{dcolumn}    
\usepackage{bm}         
\usepackage{mathtools}  
\usepackage{xcolor}     
\usepackage{microtype}  
\usepackage{orcidlink}
\usepackage{booktabs}
\usepackage{hyperref}

\newtheorem{remark}{Example}

\usepackage{hyperref}
\hypersetup{
    hidelinks
}
\begin{document}

\title{\boldmath From $Z$ to $a$: \\ High-temperature relations, subleading semi-universality, and conformal anomalies }

\preprint{TIFR/TH/26-24}

\author{Arnav Advant\orcidlink{0009-0001-1678-5573}}
\email{23b1824@iitb.ac.in}
\affiliation{Department of Physics, Indian Institute of Technology Bombay, Powai, Mumbai 400076, India}

\author{Harsh Anand\orcidlink{0009-0000-2281-3609}}
\email{harsh.anand@tifr.res.in}
\affiliation{Department of Theoretical Physics, Tata Institute of Fundamental Research, Homi Bhabha Rd, Mumbai 400005, India}

\author{Nathan Benjamin\,\orcidlink{0000-0003-3661-6563}}
\email{nathanbe@usc.edu}
\affiliation{Department of Physics and Astronomy, University of Southern California, Los Angeles, CA 90089, USA}

\author{Vipul Kumar\,\orcidlink{0009-0000-4317-608X}}
\email{vipul.kumar@tifr.res.in}
\affiliation{Department of Theoretical Physics, Tata Institute of Fundamental Research, Homi Bhabha Rd, Mumbai 400005, India}

\author{Shiraz Minwalla\,\orcidlink{0000-0002-6789-6978}}
\email{minwalla.theory@tifr.res.in}
\affiliation{Department of Theoretical Physics, Tata Institute of Fundamental Research, Homi Bhabha Rd, Mumbai 400005, India}

\author{Jyotirmoy Mukherjee\,\orcidlink{0000-0002-1822-2264}}
\email{jyotirmoy.mukherjee\_119@tifr.res.in}
\affiliation{Department of Theoretical Physics, Tata Institute of Fundamental Research, Homi Bhabha Rd, Mumbai 400005, India}

\author{Sridip Pal\,\orcidlink{0000-0002-3813-9513}}
\email{sridip@theory.tifr.res.in}
\affiliation{Department of Theoretical Physics, Tata Institute of Fundamental Research, Homi Bhabha Rd, Mumbai 400005, India}
\affiliation{Institut des Hautes \'Etudes Scientifiques (IHES), 91440 Bures-sur-Yvette, France}

\author{Asikur Rahaman\,\orcidlink{0009-0008-7814-3059}}
\email{asikur.rahaman@tifr.res.in}
\affiliation{Department of Theoretical Physics, Tata Institute of Fundamental Research, Homi Bhabha Rd, Mumbai 400005, India}

\author{Pabitra Ray\,\orcidlink{0000-0002-5553-7003}}
\email{raypabitra96@gmail.com}
\affiliation{Department of Theoretical Physics, Tata Institute of Fundamental Research, Homi Bhabha Rd, Mumbai 400005, India}

\date{September 2, 2026}

\begin{abstract}
The free energy of any CFT, $ \ln  Z(\beta; \omega_i)$, admits two expansions: high temperature ($\beta \rightarrow 0$) and fast rotation ($\omega_i \rightarrow 1$). We demonstrate that in 4$d$, locality of the thermal effective action forces $\ln Z$ to take a simple analytic form at all orders in the high temperature expansion, and further imposes an infinite number of sharp relations on the coefficients in this expansion. All are homogeneous, except at order $\beta^1$ due to the Weyl anomaly. From this, the $a$-anomaly can be extracted from the counting of operators. The relations resum in the fast-spinning expansion into differential equations in $\beta$ obeyed by the semi-universal limit and its corrections. We verify the relations in a variety of CFTs. We generalize to any even $d$, but find no similar relations at odd $d$.

\end{abstract}

\maketitle


\subsection{Introduction} 
As conformal field theories (CFT) play a key role in diverse branches of theoretical physics, from critical phenomena to quantum gravity, universal results that apply to all CFTs are of great interest.

Every CFT$_{d}$ has a spectrum of operators. The state-operator map identifies this with the spectrum of the Hilbert space of the CFT on $ S ^{d-1}$, which is efficiently encoded in the thermal partition function 
\begin{equation}\label{pfsd}
{Z}(\beta;\omega_i) :={\rm Tr}\ e^{ -\beta(D -\omega_1 J_1-\cdots -\omega_{\kappa } J_{\kappa} )},\quad  \kappa:=\left\lfloor \frac d2 \right\rfloor
\end{equation}
where $D$ is the dilatation operator and $J_i\,,~i=1,\ldots,\kappa$ commuting angular momentum operators associated with the $\kappa$ rotations in the orthogonal planes of $ R ^d$ where $ S ^{d-1}$ is embedded. As the partition function ${Z}$ has an operator-counting interpretation, it is completely scheme-independent and the central object of study in this letter. 

While the details of $Z$ are theory-dependent, its asymptotics display universality in two different limits.

The first of these is the high temperature limit. At fixed $J_i$, the energy is unbounded above, so the trace diverges for $\beta\leq 0$. Using the fact that (almost -- see later for exceptions) every consistent physical theory is well-approximated by hydrodynamics at high temperatures, the authors of \cite{Bhattacharyya_2008} argued 
that the approach to the high temperature edge $\beta\to 0^{+}$ takes the universal form
\begin{equation}\label{pfhtu}
\ln Z= \frac{\Big[ c'+ {\cal O}(\beta^2) \Big]}{\beta^{d-1}\prod_i( 1-\omega_i^2)}
,\quad \beta\to 0^+, \quad \omega_i\ \rm{fixed}\,.
\end{equation}
The dimensionless constant $c'$ is theory dependent and thus non-universal. The recent reincarnation of this fluid dynamical argument utilizes the formalism of \cite{Banerjee:2012iz, Jensen:2012jh, Shaghoulian:2015lcn} and is known as \textit{thermal effective field theory} \cite{Benjamin:2023qsc}.

The trace diverges a second time when any of $\omega_i^2$ exceeds $1$ at fixed temperature, as observed in the recent studies of Grey Galaxies \cite{Kim:2023sig, Bajaj:2024utv, Choi:2025lck} and in \cite{Kusuki:2018wpa,Benjamin:2019stq,Pal:2023cgk,Pal:2025yvz} for $2$-dimensional CFT.
The recent paper \cite{Anand:2025mfh} used thermal effective field theory to demonstrate that the approach to this divergence is also (semi)-universal. In particular,  ${Z}$
takes the `semi-universal' form 
\begin{equation} \label{sn}
\ln Z=\frac{2^{\kappa}}{\prod_i(1-\omega_i^2)}
\Big[h(\beta)+{\cal O}( 1-\omega_i^2)\Big]~,
\end{equation}
when all  $\omega_i^2$ are scaled to unity at comparable rates, i.e. in the limit
\begin{equation}\label{limstr}
\omega_i \to 1~,  ~~{\rm with}~~\beta~~{\rm and} ~~\frac{1- \omega_i^2 }{1-\omega_j^2} ~~{\rm fixed~for}~1\le i<j\le \kappa~.
\end{equation}
$h(\beta)$ in (\ref{sn}) is a theory-dependent (so non-universal) function of the inverse temperature $\beta$.
More recently, the paper 
\cite{Komargodski:2026ain} has offered a simple physical interpretation of the effective extensivity (discovered  in \cite{Anand:2025mfh}) of the entropy 
of CFTs in the limit \eqref{limstr}.

In this paper we focus on CFTs whose partition function $\ln Z$ is well described by a local thermal effective field theory that unifies \eqref{pfhtu} and \eqref{sn} at small $\beta^2$ and $x_i$, where
\begin{equation}\label{def:x}
    x_i:=(1-\omega_i^2)\,.
\end{equation}
We then demonstrate that the underlying locality forces the coefficients of these expansions to obey an infinite number of previously unanticipated universal constraints in all even dimensions.
The simplest of these  yields the striking formula \eqref{alldan} for the $a$-anomaly of every even-dimensional CFT in terms of {\it the asymptotics of operator counting} at large charges: a path from $Z$ to $a$. Intriguingly, none of these constraints have an obvious analogue in odd dimensions. In this letter we present our principal results together with a brief outline of their derivation. Additional details will appear in \cite{Advant:2026uca}. 

\subsection{Principal results}

We first focus on $d=4$ where
\begin{equation}\label{pfs}
Z(\beta;\omega_1,\omega_2)={\rm Tr} \exp(-\beta D + \beta \omega_1 J_1 + \beta \omega_2 J_2)\,. 
\end{equation}
The Boltzmann factor in $Z$ equals 
$e^{-\beta D}=e^{-\beta(E-E_{\rm Cas})}$, where $E$ is the energy and $E_{\rm Cas}$ is the Casimir energy on $S^3$.  ${Z}$ in \eqref{pfs} equals the CFT path integral on an appropriately twisted $ S^{3} \times S^1$ up to a  shift 
controlled by the $a$-anomaly coefficient (see Appendix~\ref{app:anom}). Our analysis proceeds by systematically deriving the form of the most general symmetry-allowed local thermal effective action for this path integral.
Integrating this local density over $ S^{3}$ then yields an all-orders prediction for the most general allowed structure of the partition function $\ln {Z} (\beta; x_1,x_2)$.

Our principal results are
\begin{enumerate}
\item The quantity $\beta^{3}x_1x_2\,\ln {Z}$ admits a  (generically asymptotic) power series expansion in $\beta^2$ and $x_1,x_2$  at small values of these variables. 
\item The theory-dependent coefficients that appear in this expansion are subject to an infinite number of universal linear constraints.
\item All but one of these relations are homogeneous. The exceptional relation involves the $a$-anomaly coefficient, and can be recast in the striking form\footnote{The RHS of \eqref{explrel} is real because 
$\ln Z$ is real analytic,
$(\omega_2^2)^*=\omega_1^2$,   and $\ln Z$ is symmetric in $\omega_1^2$ and $\omega_2^2$.}: \begin{equation}\label{explrel}
a=\left.\frac{3}{2} \ln  Z\left(\beta;\omega_1^2=e^{\frac{2\pi i}{3}},\omega_2^2=e^{\frac{4\pi i}{3}}\right) \right|_{\beta^1}
\end{equation}
where $\big|_{\beta^1}$ means: expand $ \ln  Z$ at small $\beta$ and retain only the term proportional to $\beta^1$.

The $a$-anomaly can thus be read off from the subleading spin weighted count of  operators at large dimension. 

\end{enumerate}
We now proceed to elaborate.

\subsubsection{The small $\beta^2, x_1,x_2 $ expansion}\label{sbo}

We demonstrate that the most general partition function ${Z}$ that follows from a local thermal effective action has the remarkably simple small $(\beta^2, x_i)$ expansion 
\begin{equation}\label{htomeg}
\ln {Z} = -\frac{1}{\beta^3 x_1 x_2}
\left[ \sum_{m, n, p=0}^\infty a_{m,n;p}x_1^m x_2^n 
\left( \beta^2 \right)^{m+n+p} \right] 
\end{equation}
where $a_{m, n;p}$ are theory-dependent dimensionless parameters with $a_{m, n;p}= a_{n, m;p}$, and the summation in the square bracket is generically expected to be asymptotic rather than convergent. 

The summations in \eqref{htomeg} can be grouped in two useful ways. First, we group all terms  with the same $\beta$ dependence to obtain 
\begin{equation} \label{htomegle} 
\ln Z=-\frac{1}{\beta^3x_1x_2}\sum_{q=0}^{\infty}S_q( x_1, x_2)\beta^{2q}   
\end{equation}
where 
\begin{equation}\label{pbq}
S_q(x_1,x_2):=\sum_{\substack{ m, n=0\\m+n \leq q}} a_{m, n;q-m-n} x_1^m x_2^n
\end{equation}
\eqref{htomegle} is the all-orders generalization of the high temperature partition function \eqref{pfhtu}.
As the summation in \eqref{pbq}
is finite, \eqref{htomegle} is accurate at small $\beta^2$,  {\it even when $x_i$ take finite values}. In fact $S_q$ is a symmetric polynomial of degree $q$ in $x_1$ and $x_2$. This analytic structure was already expected in \cite{Benjamin:2023qsc}, based on computations at low orders in the small $\beta$ expansion.

Second, we bunch together all terms of the same homogeneity in $x_1$ and $x_2$ to obtain 
\begin{equation}\label{htomegls}
\ln {Z} = -\frac{1}{\beta^3 x_1\, x_2}
\left[ \sum_{m, n=0}^\infty {h}_{m,n}(\beta)x_1^m x_2^n 
 \right] 
\end{equation}
where 
\begin{equation}\label{harel}
{h}_{m,n}(\beta) = \beta^{2m+2n} ~ \left[ \sum_{p=0}^{\infty} a_{m, n; p} \beta^{2 p} \right] 
\end{equation}
and ${h}_{m,n}(\beta)={h}_{n,m}(\beta)$ \footnote{Note that $h_{0,0}(\beta)=-4 \beta^3 h(\beta)$, see \eqref{sn}.}. 
The RHS of \eqref{harel} is a formal power series; whenever it is unambiguously Borel summable, \eqref{htomegls} applies even at finite $\beta^2$ provided $x_i$ are small, yielding a systematic all-orders improvement of the leading order semi-universal formula \eqref{sn}.

\subsubsection{Constraints on coefficients}\label{constcoeff}

We now explain that the coefficients $a_{m,n;p}$ in \eqref{htomeg} are subject to an infinite number of universal constraints. The local `thermal effective action' that computes $\ln Z^{\rm rot}$ (the Weyl-invariant part of $\ln{Z}$, see Appendix~\ref{app:anom}) is constructed out of the local data of compactification associated with the trace \eqref{pfs}. This data can be taken \cite{Jensen:2012jh} to be the metric of $S^3\times R$, together with Weyl-invariant `thermal identification' conformal Killing vector $v^\mu$ 
\begin{equation}\label{vvecmain}
v=\beta \left(\partial_\tau -i\omega_1\partial_{\phi_1} -i \omega_2\partial_{\phi_2} \right) 
\end{equation}
characterizing the twisted identification 
\begin{equation}\label{identinmain}
\begin{split}
 (\tau,  \phi_1, \phi_2) &\cong (\tau +\beta, \phi_1 - i\omega_1 \beta, \phi_2- i \omega_2 \beta)~.     
\end{split}
\end{equation}
Accounting for the Weyl transformation
of $\sqrt{g_4}$, it follows that 
\begin{equation}\label{pfformimain}
\begin{split}
    \ln {Z}^{\rm rot}&=\int  d^4x \sqrt{g_4}\frac{{\mathcal I}}{(v^2)^2} =\beta \int d\Omega_{3} \frac{{\mathcal I}}{(v^2)^2}\\
    \end{split}
\end{equation}
where ${\mathcal I}$ is a local diffeomorphism- and Weyl-invariant scalar built out of (derivatives of) $v^\mu$ and the (warped) $S^3\times S^1$ metric. Using the Weyl covariant calculus developed by Loganayagam \cite{Loganayagam:2008is}  (see Appendix~\ref{app:derivation} for a brief review), the most general ${\mathcal I}$ can be shown to be given, up to total derivatives, by polynomials (of arbitrary degree) of three distinguished Weyl-invariant scalars, ${{\rm tr} (\omega^2)}$, $v^2 {\mathcal S}$, and $v^2 {\mathcal B}_\mu (\omega^2)^{\mu\nu} {\mathcal B}_\nu$; see Appendix~\ref{app:derivation} for definitions and details. Plugging this general expansion into \eqref{pfformimain} and integrating over $S^3$, we obtain a result that takes the form \eqref{htomeg}, but is not the most general expression of this form. The requirement that $\ln Z^{\rm rot}$ is the integral of local Weyl-invariant density ${\mathcal I}$ thus, effectively,  imposes an infinite number of linear constraints on the coefficients of  \eqref{htomeg}. 

These constraints are most efficiently counted in the small $ x_i $ expansion. In \eqref{htomegls}, the number of distinct $p^{th}$ order coefficient functions, ${h}_{m, n}(\beta)={h}_{n, m}(\beta)$ with $m+n=p$, equals $\left\lfloor\frac{p}{2}+1\right\rfloor $. However the local analysis above produces no more than $\left\lfloor\frac{p}{3}+1\right\rfloor$
of these functions, establishing that the functions ${h}_{n, m}(\beta)$ obey at least $\left\lfloor\frac{p}{2}\right\rfloor -\left\lfloor\frac{p}{3}\right\rfloor$ universal linear identities at $p^{th}$ order. For $p \notin \{0,1,3\}$, we have at least one such identity at every value of $p$. The number of identities grows with $p$, asymptoting to $\frac{p}{6}$ at large $p$. These fascinating universal identities apply to  every CFT whose thermal effective action is local. 

While the infinitely many universal identities for $p \geq 4$ are all homogeneous linear differential constraints on the coefficient functions ${h}_{m,n}(\beta)$,  
the single identity at  $p=2$ is special;  it is inhomogeneous as the $d=4$ 
Weyl anomaly coefficient $a$ appears on its RHS, and takes the explicit form  
\begin{equation}\label{inhomoidentity}
\begin{aligned}
&\left(\beta\frac{d}{d\beta}-2\right)
\left(
\beta\frac{d}{d\beta}{h}_{0,0}(\beta)
+12{h}_{1,0}(\beta)
\right)
\\
&\qquad\qquad
+24{h}_{2,0}(\beta)
+24{h}_{1,1}(\beta)
=-16a\beta^4~.
\end{aligned}
\end{equation}
Upon expanding each of these linear differential constraints in a power series in $\beta^2$, one obtains an infinite number of linear constraints on the coefficient functions $a_{m, n; p}$. \footnote{We have derived these differential identities order by order in the expansion in $\beta^2$. When the functions $h_{m,n}(\beta)$ appearing in \eqref{harel} are unambiguously Borel summable, we expect \eqref{harel} to hold as a genuine functional identity for their Borel summation.
} For instance, plugging \eqref{harel} into \eqref{inhomoidentity}, expanding the LHS in a power series in $\beta^2$  and equating coefficients of $\beta^{2q}$ on both sides, one obtains the following nontrivial linear constraints  for each $q \geq 2$: 
\begin{equation}\label{relanom}
\begin{split}
&q(q-1)a_{0,0;q}
+6(q-1)a_{1,0;q-1}\\&\qquad\qquad~~~~\,
+6a_{2,0;q-2}
+6a_{1,1;q-2}
=0~~\mathrm{for}~q\ge3~,\\&a_{0,0;2}+3a_{1,0;1}+3a_{2,0;0}+3a_{1,1;0}=-2a~~\mathrm{for}~q=2~. 
\end{split}
\end{equation}
While the relations \eqref{relanom} are homogeneous for $q\geq 3$, the $a$ anomaly coefficient appears on the RHS of the constraint at $q=2$. We will explicitly check this special constraint for a variety of theories in Tab. \ref{tab:examples}.

Each of the differential identities above can be Taylor expanded in a similar manner. The resultant constraints between $a_{m,n;p}$ are easily counted in the language of the small $\beta^2$ expansion.  While the function $S_q(x_1,x_2)$ in \eqref{htomegle} and \eqref{pbq} is naively characterized by $\left\lfloor\frac{(q+2)^2}{4}\right\rfloor$ distinct coefficients, the constraints above tell us that these coefficients are actually subject to  
$\left\lfloor\frac{q^2}{12} + \frac q6 + \frac13\right\rfloor$ universal relations. The first of these occurs at $q=2$, and is simply the second equation of \eqref{relanom}. Every other constraint (at $q^{th}$ order, $q \geq 3$) is a linear and homogeneous relationship between coefficients $a_{m,n;p}$ with $p \leq q$.

Above we have obtained coefficient relations like \eqref{relanom} by Taylor expanding differential identities like \eqref{inhomoidentity}. However these relations can also be derived directly from the high temperature expansion \eqref{htomegle}.

\subsubsection{From ${Z}$ to $a$}
\label{fza}

 \eqref{relanom} tells us that the $a$-anomaly of the CFT is encoded in the high energy, spin weighted, growth of the number of operators of the CFT. In fact \eqref{relanom} may be viewed 
as the four-dimensional analog of the famous Cardy formula with one qualitative difference: while the $d=2$ Cardy formula relates the Weyl anomaly coefficient to the leading-order growth in the number of operators, \eqref{relanom} determines the $d=4$ $a$-anomaly in terms of coefficients parameterizing a subleading growth term of the partition function. 

We have already noted above that \eqref{relanom} can be recast in the elegant form \eqref{explrel}. \footnote{ The  analytic continuation  in  \eqref{explrel} is performed on the coefficients of the  $\beta^2$ expansion \eqref{htomegle}, and is unambiguous as $S_q$ in \eqref{pbq} are polynomials.}
In other words, the anomaly $a$ is encoded not by the leading $T^3=\beta^{-3}$ high temperature growth of $\ln {Z}$, but by terms that characterize the second subleading correction to this leading growth. We pause to sketch an independent path integral derivation (with a generalization to arbitrary even dimensions) of \eqref{explrel}. 

At $\omega_j^2=e^{2\pi i j/3}$, on the complexified universal cover (i.e. spacetime without thermal identification), a large conformal map $\Sigma$ acts as
\begin{equation}
\begin{split}
(\tau,\phi_1,\phi_2)\mapsto (-i\phi_2,-i\tau,\phi_1)\,,&\quad \ell_1\mapsto (1-\ell_1)^{-1}\\
    \beta\mapsto e^{-\pi i/3}\beta\,,&\quad\chi\mapsto1+e^{2\pi i/3}\chi 
\end{split}
\end{equation}
where we think of $S^3 \subset \mathbb{C}^2$ with co-ordinates $z_i$ and $|z_i|^2=\ell_i$ with $\sum_i \ell_i=1$ and $\phi_i$'s are the angular variables and $\chi=\frac{x_1\ell_1}{x_1\ell_1+x_2\ell_2}\Big|_{\omega_j^2=e^{2\pi ij/3}}$.  We can show that
\begin{equation}
\Sigma^*\left[I_q(\chi) d\chi\right]=e^{2\pi i(q+1)/3}[I_q(\chi) d\chi]\,,\quad S_q=\int_0^{1} I_q(\chi) d\chi
\end{equation}
For $q\equiv2\pmod3$, $S_q$ is invariant. The three images of $[0,1]$ form the oriented triangle, so their integrals are equal. Their sum is a closed-contour integral of a polynomial one-form $I_q(\chi) d\chi$ and hence vanishes. Thus $S_q=0$, proving 
\begin{equation}
\label{explrel1}
\ln Z\left(
\beta;
\omega_1^2=e^{\frac{2\pi i}{3}},
\omega_2^2=e^{\frac{4\pi i}{3}}
\right)
\bigg|_{\beta^{6j+1}}=0~,\qquad j\geq1~.
\end{equation}
For $q=2$, the same argument removes the Weyl-invariant part, leaving the anomaly contribution \eqref{explrel}. More details can be found in \hyperref[sec:supplemental]{\texttt{Supplemental Material}}. We remark that \eqref{explrel1} is independent of \eqref{inhomoidentity}, \eqref{relanom}.

The anomaly coefficient $a$ is famously a monotone under RG flows,  and is constant under marginal deformations \cite{Cardy:1988cwa, Komargodski:2011vj, Casini:2017vbe, Hartman:2023qdn}, so the same must be true of the RHS of \eqref{explrel}. Thus, though individual terms on the LHS of \eqref{relanom} presumably change under marginal deformation\footnote{Similar to the $\frac34$ renormalization of $a_{0,0;0}$ from weak to strong coupling in $\mathcal{N}=4$ SYM theory \cite{Gubser:1996de, Gubser:1998nz}.}, the specific combinations in \eqref{relanom} must remain constant. It would be very interesting to find a derivation of these monotonicity and non-renormalization properties directly from \eqref{relanom} or \eqref{explrel}.

\subsubsection{Illustration at ${\cal O}(\beta)$} \label{sketch}

As an illustration and check, we rederive the second of \eqref{relanom}, using the (perhaps more familiar) KK presentation \cite{Banerjee:2012iz, Bhattacharyya:2012nq, Shaghoulian:2015lcn, Benjamin:2023qsc,Benjamin:2024kdg}
of the density ${\mathcal I}$ of the thermal effective action.
 Reducing on $S^1_\beta$, the $4$-dimensional background metric becomes a $3$-dimensional metric $g_{ij}$, a Kaluza–Klein gauge field $A_i$, and a dilaton $\phi$, and Weyl invariance forces $g_{ij}$ and $\phi$ to enter only through ${\hat g}_{ij} \coloneqq e^{-2\phi}g_{ij}$, \cite{Benjamin:2023qsc, Benjamin:2024kdg} up to the anomaly:
\begin{equation}
-\ln Z^{\rm rot}(\beta, \omega_i) = \int \frac{d^{3}x  \sqrt{\hat g}}{\beta^{3}} \left[-f + c_1 \beta^2 \hat R + c_2 \beta^2 F^2 + \ldots \right],
\label{eq:thermaleft}
\end{equation}
where $f, c_1, c_2, \ldots$ are theory-dependent Wilson coefficients, and $\ln Z$ is obtained from $\ln Z^{\rm rot}$ from \eqref{summarint}.

The second of \eqref{relanom} is obtained from $\ln Z$ at order $\beta$. There are six independent four-derivative terms that contribute to \eqref{eq:thermaleft} at this order.  We list and evaluate them on the sphere in Table \ref{tab:fourderiv}. We see that
the result spans only a three-dimensional subspace of the four dimensional space of polynomials $S_2$ (see \eqref{pbq}) spanned by ($\{1, \omega_1^2 + \omega_2^2, \omega_1^4 + \omega_2^4, \omega_1^2 \omega_2^2\}$). All six terms, and so $\ln Z^{\text{rot}}\Big|_{\beta^1}$,  vanish at $\omega_k^2 = e^{2\pi i k/3}$. However in precisely the $\beta^1$ term, there is a difference between the trace $\ln Z$ \eqref{pfsd} and the $S^3 \times S^1$ path integral $\ln Z^{\rm rot}$, which is fixed by the $a$-anomaly (see Appendix~\ref{app:anom}). It follows that for any 4d CFT, we have (\ref{explrel}).

\begin{table}[t]
\centering
\renewcommand{\arraystretch}{1.7}
\begin{tabular}{cc}
\toprule
$\Theta$ &
$\displaystyle \frac{(1-\omega_1^2)(1-\omega_2^2)}{2\pi^2}
   \int d^{3}x\,\sqrt{\hat g}\;\Theta$ \\
\midrule
$\hat R^2$ &
$36 - 48\left(\omega_1^2+\omega_2^2\right)
    + \tfrac{148}{3}\left(\omega_1^4+\omega_2^4\right)
    - \tfrac{104}{3}\,\omega_1^2\omega_2^2$ \\
$\hat R_{ij}\hat R^{ij}$ & $12 - 16\left(\omega_1^2+\omega_2^2\right)
    + 20\left(\omega_1^4+\omega_2^4\right)
    - 8\,\omega_1^2\omega_2^2$ \\
$(F^2)^2$ &
$\tfrac{64}{3}\left(\omega_1^4+\omega_2^4\right)
    + \tfrac{64}{3}\,\omega_1^2\omega_2^2$ \\
$\hat R\, F^2$ &
$-24\left(\omega_1^2+\omega_2^2\right)
    + \tfrac{88}{3}\left(\omega_1^4+\omega_2^4\right)
    + \tfrac{16}{3}\,\omega_1^2\omega_2^2$  \\
$\hat R^{ij}F_{ik}F_{j}{}^{k}$ &
$-8\left(\omega_1^2+\omega_2^2\right)
    + \tfrac{40}{3}\left(\omega_1^4+\omega_2^4\right)
    + \tfrac{16}{3}\,\omega_1^2\omega_2^2$ \\
$\hat\nabla_i F_{jk}\hat\nabla^i F^{jk}$ &
$-8\left(\omega_1^2+\omega_2^2\right)
    + \tfrac{8}{3}\left(\omega_1^4+\omega_2^4\right)
    - \tfrac{16}{3}\,\omega_1^2\omega_2^2$  \\
\bottomrule
\end{tabular}
\caption{The six independent four-derivative terms in the thermal effective action, evaluated on the rotating $S^3$ background in the ensemble $\mathrm{Tr}\,e^{-\beta(D-\omega_i J_i)}$. Each contributes at order $\beta^{1}$ in the high-temperature expansion. On this background the six evaluations span only a three-dimensional subspace of the four-dimensional space of angular-velocity structures. Every term vanishes when evaluated on $\omega_1^2 = e^{\frac{2\pi i}3}, \omega_2^2 = e^{\frac{4\pi i}3}$.}
\label{tab:fourderiv} 
\end{table}

\subsection{Tests of our predictions} \label{top}

\begin{table}[h]
\centering
\renewcommand{\arraystretch}{2}
\begin{tabular}{lccccc}
\toprule
 & $a_{0,0;2}$ & $a_{1,0;1}$ & $a_{2,0;0}$ & $a_{1,1;0}$ & $a$-anomaly \\
\midrule
free scalar & $\frac{7}{360}$ & $-\frac{7}{360}$ & $\frac{1}{240}$ & $\frac{1}{144}$ & $\frac{1}{360}\ $ \\
Dirac fermion & $-\frac{7}{1440}$ & $-\frac{19}{720}$ & $\frac{7}{480}$ & $-\frac{1}{144}$ & $\frac{11}{360}\ $ \\
Maxwell & $-\frac{2}{45}$ & $\frac{2}{45}$ & $\frac{1}{120}$ & $-\frac{11}{72}$ & $\frac{31}{180}\ $ \\
holography $\times\frac{L^3}{G_5}$ & $-\frac{\pi}{16}$ & $0$ & $-\frac{\pi}{16}$ & $0$ & $\frac{\pi}{8}\ $ \\
\bottomrule
\end{tabular}
\caption{The coefficients $a_{m,n;2-m-n}$ in the notation of (\ref{pbq}). Plugging into (\ref{relanom}) gives the correct $a$-anomaly. See \hyperref[sec:supplemental]{\texttt{Supplemental Material}} for the partition functions from which $a_{m,n;2-m-n}$ can be read off.}
\label{tab:examples}
\end{table} 

In Appendix~\ref{app:test} (and further in \hyperref[sec:supplemental]{\texttt{Supplemental Material}}), we test this prediction in  ${\cal N}=4$ Yang-Mills theory at large $N$ and strong coupling, using the formulae for the classical thermodynamics of Kerr--AdS$_5$ black holes. We demonstrate 
that the key predictions \eqref{htomeg}, \eqref{inhomoidentity} (and so \eqref{explrel})  are obeyed. 
Moreover, the LHS and RHS of  \eqref{inhomoidentity} match with each other in an algebraically nontrivial manner. We view this agreement as a nontrivial check of the predictions of this paper. 

In \hyperref[sec:supplemental]{\texttt{Supplemental Material}}, we check our predictions against free scalar, fermion, and Maxwell theories in $d=4$. The free energy $\ln {Z}$ in  each of these theories can be divided into two parts: a first that results from a local thermal effective action, and a second that does not. While the first part obeys all our predictions, the second part does not take the form \eqref{htomeg} and also fails to obey the constraint equation
\eqref{inhomoidentity}. Despite this complication, \eqref{explrel} continues to hold exactly for each of these theories.

The fact that the partition functions of free theories do not quite admit the expansion \eqref{htomeg} serves to emphasize that the predictions of this paper fully apply only to theories whose thermal effective actions are local.
While free theories (i.e. theories with conserved  higher spin conserved currents, see \cite{Maldacena:2011jn, Maldacena:2012sf, Alba:2013yda, Alba:2015upa}) fail this criterion, interaction effects likely cure the relevant  non-locality (see subsection 5.3 of \cite{Anand:2025mfh}).

The assumptions of this study may also be evaded in a second class of models; those fail to restore symmetries even at arbitrarily high temperatures \cite{Komargodski:2024zmt,Hawashin:2024dpp,Chaudhuri:2026ges} and so are presumably described by the formalism of \cite{Bhattacharyya:2012xi} rather than  \cite{Banerjee:2012iz, Jensen:2012jh, Shaghoulian:2015lcn, Benjamin:2023qsc}. We leave this to future work. 

\subsection{Generalization to other dimensions}\label{higher}

The results for $d=4$ reported above admit a simple generalization to all dimensions. The partition function admits a (generically asymptotic) power series expansion that generalizes \eqref{htomeg} and \eqref{htomegls}. This power series can be reorganized into either a high-temperature expansion or a small $ x_i$ expansion.

Assuming CP invariance, the $d=4$ counting of coefficients and coefficient functions extends as well to every dimension (see Appendix~\ref{app:countingapp} below). Surprisingly enough, the results of this generalization differ qualitatively,  depending on whether $d$ is even or odd. 

In every even dimension, the  counting of Appendix~\ref{app:countingapp} demonstrates that the locality of the thermal effective action produces  an infinite number of constraints on the coefficients of the symmetric polynomials $S_q$ in the high-temperature expansion, and also on the coefficient functions that appear in the small $ x_i $ expansion. In odd dimensions, on the other hand, these constraints simply disappear (at least our methods uncover no such constraints).  It would be interesting to physically understand  this stark difference.

Finally, we generalize \eqref{explrel} to arbitrary even $d$ including $d=2$. The $a$-anomaly coefficient is given in terms of the thermal partition function by the striking formula
\begin{equation}\label{alldan}
a_{2\kappa}
=
(-1)^{\kappa}\left.
\frac{(\kappa+1)}
{\Gamma\!\left(\kappa+1\right)}
 \ln  {Z}\!\left(
\beta;\,
\omega_j^2=e^{\frac{2\pi i j}{\kappa+1}}
\right)
\right|_{\beta^1}   
\end{equation}
where the symbol $\left.\right|_{\beta^1}$ 
was defined under \eqref{explrel}.\footnote{There are various conventions for the $a$-anomaly in $d$-dimensions; our definition for $a_d$ follows Appendix C of \cite{Benjamin:2023qsc}. } See \hyperref[sec:supplemental]{\texttt{Supplemental Material}} for a sketch of a proof of \eqref{alldan}, which is a generalization of the proof in $d=4$ to arbitrary $d$.

\subsection{Discussion}

The study of this paper establishes the simple analytic structure \eqref{htomeg}, \eqref{htomegle}, \eqref{pbq}, \eqref{htomegls}, \eqref{harel} for the rotating thermal partition function as a function of $\beta^2$ and $x_i$, and uncovers several universal relations between coefficients, including a formula for the $a$ anomaly.

It would be instructive to find a direct path-integral derivation of all the constraints enumerated in this paper, either along the lines of \eqref{explrel1}, or 
perhaps using ideas similar to those in \cite{Jensen:2012kj, Jensen:2013kka}. Such an exercise might expose a link between (\ref{explrel}) and the $a$-theorem. 

\eqref{explrel} is plausibly related to the well established link between anomaly coefficients and the growth of superconformal indices \cite{Kinney:2005ej, Bhattacharya:2008zy} in the so-called Cardy limit (see e.g. \cite{DiPietro:2014bca,ArabiArdehali:2014otj,ArabiArdehali:2014cey, ArabiArdehali:2015iow,ArabiArdehali:2015ybk,Choi:2018hmj, Kim:2019yrz, Cabo-Bizet:2019osg, Gadde:2020bov, Goldstein:2020yvj, Jejjala:2021hlt, Cassani:2021fyv,Ardehali:2021irq, Ohmori:2021dzb,Jejjala:2022lrm}).

In 4d, the CPT theorem forces the partition function to obey \begin{equation}
    Z(\beta;\omega_1,\omega_2) = Z(\beta;-\omega_1,\omega_2)~,
\end{equation}
even for CFTs that violate parity (or CP) symmetry.\footnote{This can be seen by conjugating the trace (\ref{pfsd}) by the CPT operator. A related fact is that in 4d, CPT forbids a pure Chern-Simons term involving only the graviphoton from showing up in the thermal effective action (\ref{eq:thermaleft}) \cite{Banerjee:2012iz, Jensen:2013rga}. We are very grateful to Yifan Wang for relevant discussions.} However, for CFTs in $d \equiv 2~(\text{mod }4)$ dimensions, there is no such identity, and parity violation is visible in the partition function. The analog of our constraints for parity-violating 6d CFTs is a separate problem.

One can also extend the analysis to the asymptotics of OPE coefficients \cite{Buric:2025uqt,Simmons-Duffin:2025qox,Benjamin:2023qsc,Barrat:2025nvu,Barrat:2025twb,Barrat:2025wbi,Iliesiu:2018fao,Buric:2024kxo,Buric:2025fye,Buric:2026pes}, to the study of theories with additional conserved charges (and possibly critical chemical potentials \cite{Choi:2024xnv}), and away from equilibrium,  (perhaps in a holographic context generalizing \cite{Bhattacharyya:2007vjd, Bhattacharyya:2008xc, Bhattacharyya:2008mz, Bhattacharya:2008zy}), and to squashed sphere partition functions \cite{Parmentier:2026aqh}.

It would be interesting to investigate the complex analytic structure of $\ln Z$,  which appears to have a dense set of  singularities \cite{Benjamin:2024kdg}. One practical strategy might be to generalize exact resummation of the small $\beta$ expansion of the free scalar theory \cite{Lei:2024oij} to include fermions and gauge fields, and then to track the evolution (nonperturbatively well understood but so far very special) analytic structure of free theories to first nontrivial order in the coupling $\lambda$ in  ${\cal N}=4$ Yang-Mills theory. 

As the  small $\beta^2$ and $x_i$ behavior of the torus partition function of a 2d CFT is easily extracted from modular invariance, it is tempting to view the expansion \eqref{htomeg} (together with the universal constraints) as a higher dimensional version of modular invariance, and explore the analog of the two-dimensional modular bootstrap program \cite{Hellerman:2009bu}, by imposing twin requirements of \eqref{htomeg} (plus constraints) at small $\beta$,  and integrality of coefficients \cite{Kaidi:2020ecu,Fitzpatrick:2023lvh,Chiang:2023qgo,Benjamin:2026lbj} at large $\beta$.

\subsection{Acknowledgments}
We thank J. Choi, E. Lee, O. Parrikar, P. Rath, S. Trivedi, and especially A. Gadde and Y. Wang for very useful discussions. We would also like to thank K. Bajaj, J. Bhattacharya, J. Choi, T. Hartman, K. Jensen, S. Kim, Z. Komargodski, R. Loganayagam, C. Patel, J. Penedones, L. Rastelli, V. Schomerus, D. Simmons-Duffin, S. van Leuven, and S. Wadia for their comments on a preliminary version of this manuscript. The direct argument for the formula \eqref{alldan}, 
presented in the last subsection of \hyperref[anomproof]{\texttt{Supplemental Material}}, was developed in conversation with Claude Fable 5 (Anthropic). The authors specified the setup and the target statement,
checked and verified each step, and the final result. N.B. is supported by the U.S. Department of Energy, Office of Science, under grant Contract Number DE-SC0026324. The work of A.A., H.A., V.K., S.M., J.M., S.P., A.R. and P.R. was supported by the Department of Atomic Energy, Government of India, under Project Identification Number RTI-4012 and the Infosys Endowment for the study of the Quantum Structure of Spacetime. A.A., H.A., V.K., S.M., J.M., S.P., A.R. and P.R. would also like to acknowledge their debt to the
people of India for their steady support of the study of basic science.

\bibliography{biblio}

\clearpage

\appendix

\refstepcounter{section}
\label{app:anom}
\section{Appendix A: Partition functions and the Weyl anomaly}

Let $Z^{\rm round}$ denote the Euclidean path integral of a four-dimensional CFT on the twisted $S^3 \times S^1$ that computes the trace \eqref{pfs}. This path integral counts states on a round $S^3$ graded by energy and angular momentum. As the energy of each state exceeds the dimension of its corresponding operator by the Casimir energy \footnote{See \cite{Herzog:2013ed} for this value in general $d$ (even). This is computed in the conventions that set $S_{\text{ct}}$ in (2.21) of \cite{Benjamin:2023qsc} to zero.} it follows that
\begin{equation}\label{roundhat}
-\ln Z^{\rm round}= -\ln {Z} + \beta\frac{3 a}{4}~.
\end{equation}
The twisted $S^3 \times S^1$ partition function can also be evaluated in a Weyl frame in which the usual metric on $S^3 \times S^1$ is rescaled by a factor $\frac{\beta^2}{v^2}$, ($v^\mu$ is defined in \eqref{vvecmain}). As the CKV  $v^\mu$ is Weyl invariant, the same is true of the rescaled metric  $\frac{\beta^2 g_{\mu\nu}}{v^2}= 
\frac{\beta^2 g_{\mu\nu}}{v^\mu g_{\mu\nu} v^\nu}$, and hence also the partition function, $Z^{\rm rot}$ \cite{Benjamin:2023qsc}. It follows from the usual formulae (see e.g. \cite{Eling:2013bj, Benjamin:2023qsc}) for the Weyl transformations of partition functions that 
\begin{equation}\label{wt} 
-\ln Z^{\rm round}= -\ln Z^{\rm rot} -\beta\frac{a}{12}
\frac{ (x_1-x_2)^2}
{x_1x_2
}~.
\end{equation}
Combining \eqref{roundhat}
and
\eqref{wt}
we find 
\begin{equation}\label{summarint}
 -\ln {Z}=  -\ln Z^{\rm rot} - \beta \frac{a}{12}
\left( \frac{x_1}{x_2} + \frac{x_2}{x_1} + 7\right)
\end{equation}
where $\ln Z^{\rm rot}$ is Weyl invariant. 

\refstepcounter{section}
\label{app:derivation}
\section{Appendix B: The local structure of ${\ln Z}^{\rm rot}$ }

In our construction of the most general Weyl invariant scalar density ${\mathcal I}$, it is convenient to utilize the Weyl invariant calculus developed by Loganayagam in \cite{Loganayagam:2008is}. The formalism of \cite{Loganayagam:2008is} requires the specification of a distinguished  Weyl invariant vector;  in our work we choose  this to be the conformal Killing vector $v^\mu$ \eqref{vvecmain}. We define a 
tensor $Q^{\lambda\cdots}_{\quad\,\nu\cdots}$ to have 
Weyl weight $w$ if the Weyl transformation ${\tilde g}_{\mu \nu} = e^{2 \phi} g_{\mu \nu}$ induces the transformation ${\tilde Q}^{\lambda\cdots}_{\quad\,\nu\cdots}=e^{-w \phi} Q^{\lambda\cdots}_{\quad\,\nu\cdots}$. Note that the metric $g_{\mu\nu}$ itself has  Weyl weight $-2$. The Weyl covariant derivative \footnote{Our conventions are related to Loganayagam's conventions  \cite{Loganayagam:2008is} by $\mathcal{A}_\mu=\nabla_\mu\sigma$. We can see that under Weyl rescaling, $\tilde{\sigma}=\sigma+\phi$ and $\tilde{\mathcal{A}}_\mu=\mathcal{A}_\mu+\nabla_\mu\phi$. } of a tensor of Weyl weight $w$  is given in terms of 
\begin{equation}\label{betadef}
\sigma:= \frac{1}{2}\ln v^2~, ~~~
\end{equation}
by
\begin{equation}\label{wcderr}
\begin{split}
D_\alpha Q^{\mu\cdots}_{\quad\,\nu\cdots}&=\nabla_\alpha Q^{\mu\cdots}_{\quad\,\nu\cdots}+w\nabla_\alpha\sigma Q^{\mu\cdots}_{\quad\,\nu\cdots}\\&+\big[g_{\alpha\lambda}\nabla^\mu\sigma-\delta^\mu_{~\lambda}\nabla_\alpha\sigma-\delta^\mu_{~\alpha}\nabla_\lambda\sigma\big]Q^{\lambda\cdots}_{\quad\,\nu\cdots}+\cdots\\&-\big[g_{\alpha\nu}\nabla^\lambda\sigma-\delta^\lambda_{~\alpha}\nabla_\nu\sigma-\delta^\lambda_{~\nu}\nabla_\alpha\sigma\big]Q^{\mu\cdots}_{\quad\,\lambda\cdots}-\cdots~.
\end{split}
\end{equation}
Here $\nabla_\alpha$ is the ordinary covariant derivative. This definition is useful because the LHS of \eqref{wcderr} can be verified to transform under Weyl transformations in exactly the same way as  $Q^{\mu\cdots}_{\quad\,\nu\cdots}$ itself.  The derivative \eqref{wcderr} obeys the Leibniz rule and annihilates the metric tensor. The commutator of two Weyl covariant derivatives can be used to define the Weyl covariant curvature tensor ${\cal R}_{\alpha \beta\mu}^{~~~~\,\lambda}$ via 
\begin{equation}\label{wcder1}
[D_\alpha, D_\beta] V_\mu 
={\cal R}_{\alpha \beta\mu}^{~~~~\,\lambda} V_\lambda
\end{equation}
where $V_\mu$ is an arbitrary one-form. One then proceeds to define the Weyl covariant Schouten tensor ${\mathcal S}_{\mu\nu}$ via 
\begin{equation}\label{schouten}
    \mathcal{S}_{\mu\nu}=\frac{1}{(d-2)}\left(\mathcal{R}_{\mu\nu}-\frac{\mathcal{R}g_{\mu\nu}}{2(d-1)}\right), 
\end{equation}
its trace ${\mathcal S} = \mathcal{S}^{\mu}_{~\mu}$, and the vector ${\mathcal B}_{\mu}= {\mathcal S}_{\mu\nu}v^\nu $. 

We now apply this formalism to the situation at hand. Since $S^3 \times R$ is conformally flat, its Weyl covariantized (equals ordinary) Weyl tensor vanishes. This condition determines ${\cal R}_{\alpha \beta\mu}^{~~~~\,\lambda}$ in terms of  $\mathcal{S}_{\mu\nu}$ as 
\begin{equation}\label{wcwtimp}
\mathcal{R}_{\mu\nu\rho\sigma} = -\left(
g_{\nu\rho} \mathcal{S}_{\mu\sigma}
- g_{\mu\rho} \mathcal{S}_{\nu\sigma}
- g_{\nu\sigma} \mathcal{S}_{\mu\rho}
+ g_{\mu\sigma} \mathcal{S}_{\nu\rho}\right)~. 
\end{equation}
As $v^\mu$ is a CKV, it obeys the `conformal Killing lemma'
\begin{equation}\label{wcder2}
D_\alpha D_\beta v_\mu=
{\cal R}_{\mu\beta\alpha}^{~~~~\,\lambda} v_\lambda~.
\end{equation}
Upon taking the inner product of \eqref{wcder2} with $v^\beta$ 
we obtain an equation that allows us to express $\mathcal{S}_{\mu\nu}$ (hence ${\cal R}_{\mu\beta\alpha}^{~~~~\,\lambda}$) in terms of ${\mathcal S}$,  ${\rm tr} (\omega^2)$ and ${\mathcal B}_\mu$
\begin{equation}\label{schouen_exp}
v^2\mathcal{S}_{\mu\nu} = -\frac{1}{4} (\omega^2)_{\mu\nu} + v_{\mu} \mathcal{B}_{\nu}+v_{\nu} \mathcal{B}_{\mu} + \frac{g_{\mu\nu}}{2} \left(v^2 \mathcal{S}+\frac{\operatorname{tr}(\omega^2)}{4}   \right)~.
\end{equation}
We are also able to demonstrate that the Weyl covariantized Lie derivative of every relevant tensor vanishes in the $v^\mu$ direction. This fact, together with the vanishing of the Weyl covariantized Cotton tensor (which establishes that $D_\alpha {\cal S}_{\mu\nu}$ is completely symmetric in all indices) allows us to demonstrate that all Weyl covariant derivatives of  ${\mathcal B}_\nu$, and ${\cal S}$ can both be rewritten as polynomial expressions in 
${\mathcal S}$,  ${\rm tr} (\omega^2)$ and ${\mathcal B}_\mu$. 
Repeatedly (Weyl) differentiating \eqref{wcder2} then establishes that the same is true for all Weyl covariant derivatives of $v^\mu$. We thus arrive at the following key theorem: every local Weyl invariant tensor (that can be constructed out of the metric on $S^3 \times R$ plus the CKV $v^\mu$) can be written as a sum of products of the quantities
$\dfrac{g_{\mu\nu}}{v^2}$,
$v^2 g^{\mu\nu}$,
$v^\mu$,
$\omega_{\mu}{}^{\nu}$,
$v^2 \mathcal{S}$, and
$\mathcal{B}_{\mu}$,
where $\omega_{\mu \nu}= {D}_\mu v_\nu -{D}_\nu v_\mu$.  This theorem applies whenever the $ d$-dimensional metric is conformally flat (as is the case for $S^{d-1} \times R$) and 
the Weyl invariant vector $v^\mu$ is a CKV of this manifold (as is the case for $v^\mu$ defined in \eqref{vvecmain}). (See \hyperref[sec:supplemental]{\texttt{Supplemental Material}} for more details on the classification.)

The theorem quoted above can now be used to parameterize the most general diffeomorphism, Weyl and  CPT invariant form of the density function ${\mathcal I}$. In $d=4$, it is easily shown that this is given by sums of products of ${\rm tr}(\omega^2)$, $v^2{\mathcal S}$, $v^2{\mathcal B}^2$ and $v^2 {\mathcal B}_\mu (\omega^2)^{\mu\nu} {\mathcal B}_\nu$. (Using that $v^\mu$ and $g_{\mu\nu}$ are both CPT even, while $\partial_\mu$ is CPT odd, it follows that all scalars built out of a single $\epsilon_{\mu\nu\rho
\beta}$ (and the data above) are necessarily CPT odd, and so cannot appear in $\ln Z$. This conclusion holds in  every $d\equiv 0~(\text{mod }4)$, but not in $d\equiv 2~(\text{mod }4)$ dimensions (e.g. $\omega\wedge\omega\wedge v\wedge{\mathcal B}$ is CPT even in $d=6$).)
It is also possible to demonstrate that any local expression involving one or more factors of  $v^2{\mathcal B}^2$ can be re-expressed in terms of other scalar quantities independent of $v^2{\mathcal B}^2$  plus a total derivative. The results \eqref{htomeg}, \eqref{inhomoidentity}, \eqref{relanom}, and the all orders counting presented above \eqref{inhomoidentity} all follow 
upon integrating over $S^3$.

\refstepcounter{section}
\label{app:countingapp}
\section{Appendix C: Counting}

Generalizing to $d$ dimensions, one can establish the following results. The number of independent functions of temperature that appear at $p^{th}$ order in the $d$-dimensional generalization of \eqref{htomegls} equals the coefficient of $z^p$ in 
\begin{equation}\label{countingsympol}
\frac{1}{(1-z)(1-z^2) \ldots \left(1-z^{\left\lfloor\frac{d}{2}\right\rfloor}\right)}~.
\end{equation}
However the number of independent local densities ${\mathcal I}$ is no larger than the coefficient of $z^p$ in  
\begin{equation}\label{countinglocal}
\frac{1}{(1-z)(1-z^2) \ldots \left(1-z^{\left\lfloor\frac{d-1}{2}\right\rfloor}\right) \left(1-z^{\left\lfloor\frac{d-1}{2}\right\rfloor +2}\right)}~.
\end{equation}

The coefficient of $z^p$ in \eqref{countinglocal} is $\leq$ (and generically less than) the corresponding coefficient 
in \eqref{countingsympol} when $d$ is even, but the inequalities are reversed when $d$ is odd. Consequently, the coefficient functions of the semi-universal expansion obey an infinite number of universal relations in even $d$ (the number of relations scales as $\sim\frac{p^{\kappa-1}}{(\kappa+1)! (\kappa-1)!}$ at large $p$). However the counting above yields no analogous relations in odd $d$. 

In a similar manner, the number of independent coefficients at order $\beta^{2q}$, in the general $d$ version of \eqref{pbq}, 
is given by the coefficient of $z^q$ in 
\begin{equation}
\label{countingsympolder}
\frac{1}{(1-z)^2(1-z^2) \ldots \left(1-z^{\left\lfloor\frac{d}{2}\right\rfloor}\right)}~.
\end{equation}
On the other hand, the coefficients that characterize the vector space of polynomials $S_q$ that arise from a local density ${\cal I}$ are no more\footnote{In fact, we believe that this  equals the coefficient of $z^q$ in \eqref{countinglocalder} when $d$ is even, and the coefficient of $z^q$ in \eqref{countingsympolder} when $d$ is odd. We have checked this expectation at low orders in several different dimensions.} than the coefficient of $z^q$ in 
\begin{equation}\label{countinglocalder}
\frac{1}{(1-z)^2(1-z^2) \ldots \left(1-z^{\left\lfloor\frac{d-1}{2}\right\rfloor}\right) \left(1-z^{\left\lfloor\frac{d-1}{2}\right\rfloor +2}\right)}~.
\end{equation}

The second coefficient is generically smaller than the first 
in even dimensions (the difference equals $\frac{q^\kappa}{\kappa!(\kappa+1)!}$ at large $q$). However, the inequalities are reversed in odd $d$.

\refstepcounter{section}
\label{app:test}
\section{Appendix D: Tests}

In the limit specified in \eqref{limstr}, the Kerr--AdS$_5$ partition function takes the form
\begin{equation}
\begin{split}
\ln {Z}_{\rm Kerr-AdS_5}
&=
-\frac{1}{\beta^3x_1x_2}
\Big[
h_{0,0}
+ h_{1,0}\big(x_1+x_2\big)
\\&\!+ h_{2,0}\big(x_1^2+x_2^2\big)
+ h_{1,1}x_1x_2
\Big]
+{\cal O}(x_i)
\label{kerrexp}\end{split}
\end{equation}
with,
\begin{equation}
\begin{split}
{h}_{0,0}(r_0)&=-\frac{2\pi^5 r_0^2(r_0^2-1)^3}
{G_5(2r_0^2-1)^4}~,\\
{h}_{1,0}(r_0)&={h}_{0,1}(r_0)=\frac{2\pi^5 r_0^4(r_0^2-1)}
{G_5(2r_0^2-1)^4}~,\\
{h}_{2,0}(r_0)&={h}_{0,2}(r_0)=-r_0^2\, h_{1,1}(r_0)~,\\  h_{1,1}(r_0)&=\frac{2\pi^5 r_0^6}
{G_5(r_0^2-1)(2r_0^2-1)^4(2r_0^2+1)}~.   
\end{split}
\end{equation}
Here $\ln {Z}_{\rm Kerr-AdS_5}$  is the trace defined in \eqref{pfs}, whose relation to other definitions $\ln {Z}^{\rm round}_{\rm Kerr-AdS_5}$ and  $\ln {Z}^{\rm rot}_{\rm Kerr-AdS_5}$, commonly employed in studies of holography, is given in Appendix~\ref{app:anom}.
The relation between $r_0$ and the inverse temperature $\beta$ is
\begin{equation}
\beta=\frac{2\pi r_0}{2r_0^2-1}~,
\qquad
\beta\frac{d}{d\beta}
=
-\frac{r_0(2r_0^2-1)}
{2r_0^2+1}\frac{d}{dr_0}
\label{eq:r0-beta}
\end{equation}
which, upon substitution into \eqref{inhomoidentity}, indeed satisfies the constraint with the Weyl-anomaly coefficient of the dual four-dimensional CFT given by
\begin{equation}\label{aholo}
a=\frac{\pi}{8G_5} =\frac{N^2}{4}
\end{equation}
Here we have set the AdS$_5$ curvature radius to $L=1$ and used the relation $\frac{1}{G_5}=\frac{2 N^2}{\pi}$. 
\footnote{Note that the Kerr Black hole studied in this Appendix dominates the ensemble only for $\beta< 2\pi$. The Kerr black hole develops a superradiant instability at this critical value of $\beta$ \cite{Anand:2025mfh, Kim:2023sig, Bajaj:2024utv}.}

\clearpage
\onecolumngrid

\setcounter{equation}{0}
\renewcommand{\theequation}{S\arabic{equation}}

\section*{Supplemental Material}
\label{sec:supplemental}

\subsection{Local data on a conformally flat space with a CKV}

\subsubsection{The conformal Killing vector lemma and the vanishing of the Cotton tensor}

In Appendix \ref{app:derivation} we have explained that the 
equation \eqref{schouen_exp} and the complete symmetry 
of $D_\alpha {\mathcal S}_{\mu\nu}$ together play a key role in the enumeration of local data. In this subsubsection we will give a simple proof of these two equations. 

Our proof proceeds by first working in a Weyl frame in which the metric $g_{\mu\nu}$ of our ($S^{d-1} \times S^1$) manifold has been rescaled by the factor $\frac{\beta^2}{v^2}$. 
\footnote{This is the same as the Weyl frame in which the partition function, $Z^{\rm rot}$ was computed (see Appendix \ref{app:anom} for details).}

In this Weyl frame  $\tilde{v}^2=  \frac{\beta^2}{v^2} v^\alpha g_{\alpha \beta} v^\beta= \beta^2$, and so is a constant.  It is easy to check that any conformal Killing vector $v^\mu$ whose norm is 
nonzero and constant is necessarily a Killing vector \footnote{One proof of this goes as follows. Dot both indices of the CKV equation with $v$. The constancy of the $v^2$ then tells us that the LHS vanishes. The RHS, on the other hand, is proportional to $v^2 \times \nabla . v$. Since $v^2$ is nonzero, it follows that $\nabla . v$ vanishes, and so the CKV is, in fact, a Killing vector.}. In this Weyl frame, therefore, $v_\alpha$ obeys 
\begin{equation}\label{killequ}
\nabla_\alpha v_\beta +\nabla_\beta v_\alpha=0~.
\end{equation}

Now, as $v^2=\beta^2$ is a constant, it follows that Loganayagam's gauge field ${\mathcal A}_\mu$  vanishes in this Weyl frame, so that the covariant derivative $D_\alpha$ \eqref{wcderr} reduces to the ordinary covariant derivative $\nabla_\alpha$.
It follows that in this frame, \eqref{killequ} can be rewritten as
\begin{equation}\label{killequcov}
D_\alpha v_\beta +D_\beta v_\alpha=0~.
\end{equation}
As \eqref{killequcov} is Weyl covariant, it applies in every frame.

The conformal Killing vector lemma \eqref{wcder2} can now be derived as follows. In ordinary geometry, Killing vectors famously obey the equation
\begin{equation}\label{kvl}
\nabla_\alpha \nabla_\beta v_\mu=
{R}_{\mu\beta\alpha}^{~~~~\,\lambda} v_\lambda~.
\end{equation}
In our special frame, it follows from the vanishing of ${\mathcal A}_\mu$ that \eqref{kvl}
can equivalently be written as 
\begin{equation}\label{kvlcov}
D_\alpha D_\beta v_\mu=
{\mathcal R}_{\mu\beta\alpha}^{~~~~\,\lambda} v_\lambda~.
\end{equation}
Since \eqref{kvlcov} is covariant, it applies in every frame.

The symmetry of $D_{\alpha} {\mathcal S}_{\mu\nu}$ can be argued in a very similar manner. The \textit{Codazzi identity} asserts that the Cotton tensor of a manifold vanishes if its Weyl tensor vanishes\footnote{This holds for $d \geq 4$. For $d = 3$, the Weyl tensor vanishes identically and the vanishing of the Cotton tensor becomes the defining condition of conformal flatness.}. As the latter condition is met in our case, it follows that 
\begin{equation}\label{vancot}
\nabla_\alpha { S}_{\mu \nu}=\nabla_{\mu} { S}_{\alpha \nu}
\end{equation}
where $S_{\mu\nu}$ is the ordinary Schouten tensor. The vanishing of ${\mathcal A}_\mu$ in our Weyl frame allows us to rewrite \eqref{vancot} as 
\begin{equation}\label{vancotcov}
D_\alpha {\mathcal S}_{\mu \nu}=D_{\mu} {\mathcal S}_{\alpha \nu}~.
\end{equation}
(The Weyl covariant Schouten tensor ${\mathcal S}_{\mu\nu}$ equals the usual Schouten tensor $S_{\mu\nu}$ in this special frame.) However, as \eqref{vancotcov} is covariant, it holds in every Weyl frame. 

\subsubsection{Construction of the basis}

The Killing vector lemma \eqref{wcder2} allows us to write higher derivatives of the velocity in terms of the derivatives of the Schouten tensor. \eqref{schouen_exp} then tells us that this task is further reduced to finding derivatives of the vector $\mathcal{B}_\mu$ and $\mathcal{S}$.

Repeated use of \eqref{wcder1}, \eqref{vancotcov} and \eqref{schouen_exp} along with the fact that the Lie derivative along $v^\mu$ vanishes for any tensor allows us to show that derivatives of $\mathcal{B}_\mu$ and $\mathcal{S}$ can be rewritten as polynomials of the base quantities. With this analysis, any tensor of such a manifold can be written as polynomials of the following quantities
\begin{equation}\label{supp_tlist} 
  \begin{tabular}{||c|c|c||}
  \hline
   Quantity  & Weyl-weight &  $\beta$ scaling \\
   \hline 
    $v_\mu$ & $-2$  & $1$  \\
    $g_{\mu\nu}$ & $-2$  & $0$\\
    $\omega_{\mu\nu}$ & $-2$  & $1$\\
    $\mathcal{S}$ & $2$  & $0$\\
     $\mathcal{B}_{\mu}$ & $0$  & $1$\\  
    \hline
\end{tabular}  
\end{equation}
\subsubsection{Local scalars}
The case of scalars has another notable property, which is responsible for the infinite number of constraints between the coefficient functions. Note that \eqref{supp_tlist} tells us that the local scalars of the theory in $d$ dimensions are
\begin{equation}\label{locstr}
    v^2,\quad\mathcal{S},\quad \mathcal{B}^2,\quad{\rm tr}(\omega^{2k}),\quad\mathcal{B}_\mu (\omega^{2k})^{\mu\nu}\mathcal{B}_\nu, \quad k = 0,1,2... \left\lfloor \frac{d-1}{2} \right\rfloor~.
\end{equation}
The Cayley-Hamilton theorem allows us to write higher powers of $\omega_{\mu\nu}$ in terms of linear combinations of lower powers. It can be shown using \eqref{wcder1}, \eqref{vancotcov}, and \eqref{schouen_exp} that the divergence of a vector $ (\omega^{2(k+1)})^{\mu\nu}\mathcal{B}_\nu$,  with $k=0 \ldots \left\lfloor \frac{d-3}{2} \right\rfloor$, can be re-written in terms of scalars, one of which is $\mathcal{B}_\mu (\omega^{2k})^{\mu\nu}\mathcal{B}_\nu$. In 4d, there is only one such equation (corresponding to $k=0$). This equation takes the form
\begin{equation}
    4\mathcal{B}^2 =-D_\mu(\omega^{\mu\nu}\mathcal{B}_\nu)+  v^2\mathcal{S}^2~,\qquad \omega_\mu^{~\nu}\mathcal{B}_\nu=-\frac{v^2}{3}D_\mu\mathcal{S}~.
\end{equation}
So every time we have a scalar of the form $\mathcal{B}_\mu (\omega^{2k})^{\mu\nu}\mathcal{B}_\nu$ with $k=0 \ldots \left\lfloor \frac{d-3}{2} \right\rfloor$, we can re-write it (up to total derivatives) in terms of polynomials of the other local scalars (i.e. in terms of polynomials that omit this quantity). It is clear we can keep repeating this process until we reach the top of the ladder. As the partition function is given by an integral over local scalars, the density is a polynomial made from
\begin{equation}\label{localstr2}
    \quad v^2\mathcal{S}, \quad v^2\mathcal{B}_\mu (\omega^{2\lfloor \frac{d-1}2\rfloor})^{\mu\nu}\mathcal{B}_\nu, \quad  {\rm tr}(\omega^{2k})~, \quad k = 0,1,2... \left\lfloor \frac{d-1}{2} \right\rfloor~.
\end{equation}
The scaling with $\beta^2$ of these scalars can be read off 
from \eqref{supp_tlist}. The scaling with $(1-\omega_i^2)$ is a bit more involved; instead of using the quantities ${\rm tr}(\omega^{2k})$ themselves, one uses ${\rm tr} (\omega^2)$, 
${\rm tr} (\omega^4) -c \left( {\rm tr} (\omega^2) \right)^2$ (with $c$ chosen to minimize the scaling with $(1-\omega_i^2)$) and so on. For example, in $d=6$
the good basis consists of $\mathrm{tr}(\omega^2)$, $\left[\mathrm{tr}(\omega^4)-\frac{1}{4}\left(\mathrm{tr}(\omega^2)\right)^2\right]$. 

In a similar manner, one also replaces $(\omega^{2k})^{\mu\nu}$ in $(\omega^{2k})^{\mu\nu}\mathcal{B}_\nu $, chosen so that this quantity scales in the slowest possible manner with $(1-\omega_i^2)$. For instance, in $d=6$
one finds that this combination is given by
$v^2\mathcal{B}_\mu \left[(\omega^4)^{\mu\nu}-\frac{1}{4}\mathrm{tr}(\omega^2)(\omega^2)^{\mu\nu}\right]\mathcal{B}_\nu$.

With this understanding, we find that the $(1-\omega_i^2)$ scaling of  each of the scalars
matches its scaling with $\beta^2$, with the sole exception of ${\rm tr}(\omega^2)$, which is of order $\beta^2$, even though it is of order $(1-\omega_i^2)^0$. These charge assignments yield the counting presented in Appendix \ref{app:countingapp}.
Thus in $d=6$ the useful version of \eqref{locstr} is 
\begin{equation}\label{fullbasis}
{\rm tr}(\omega^2),\qquad v^2\mathcal{S}, \qquad\left[\mathrm{tr}(\omega^4)-\frac{1}{4}\left(\mathrm{tr}(\omega^2)\right)^2\right],\qquad v^2\mathcal{B}_\mu(\omega^2)^{\mu\nu}\mathcal{B}_\nu,\qquad v^2\mathcal{B}_\mu \left[(\omega^4)^{\mu\nu}-\frac{1}{4}\mathrm{tr}(\omega^2)(\omega^2)^{\mu\nu}\right]\mathcal{B}_\nu~.
\end{equation} 
These scalars scale respectively like $[\beta^2,(1-\omega_i^2)^0]$, $[\beta^2, (1-\omega_i^2)]$, 
$[\beta^4, (1-\omega_i^2)^2 ]$, $[\beta^6, (1-\omega_i^2)^3 ]$ and $[\beta^8,(1-\omega_i^2)^4]$. In a similar manner, the accurate $d=6$
version of \eqref{localstr2} is 
\begin{equation}\label{fullglobbasis}
{\rm tr}(\omega^2),\qquad v^2\mathcal{S}, \qquad\left[\mathrm{tr}(\omega^4)-\frac{1}{4}\left(\mathrm{tr}(\omega^2)\right)^2\right],\qquad v^2\mathcal{B}_\mu \left[(\omega^4)^{\mu\nu}-\frac{1}{4}\mathrm{tr}(\omega^2)(\omega^2)^{\mu\nu}\right]\mathcal{B}_\nu~.
\end{equation}

Similar constructions work in arbitrary dimensions, and underlie the counting presented in Appendix \ref{app:countingapp}.

\subsubsection{d = 4}

From the analysis above, the 4d `Weyl-invariant' partition function can be written as 
\begin{equation}
    \ln{Z}^{\rm rot} = \int d^4x \sqrt{g_4}\frac{1}{(v^2)^2} \left(\sum_{m,n=0}^\infty  {\mathcal C}_{m,n}\left(\frac{1}{8}{\rm tr}(\omega^2)\right) ~\left(v^2 \mathcal{S}\right)^{m} \left( {v^2\mathcal{B}_\mu(\omega^2)^{\mu\nu}\mathcal{B}_\nu}\right)^{ n}\right)~.
\end{equation}
After explicit computation, we find that the $\nu = \mathcal{O}(1  -\omega_i^2)$ and $\beta^2$ scalings of the relevant scalars are
\begin{equation}\label{supp_tlist2} 
  \begin{tabular}{||c|c|c||}
  \hline
   Quantity  & $\beta^2$ scaling &  $\mathcal{O(\nu)}$ \\
   \hline 
    $v^2$ & $1$  & $\nu$  \\
    ${\rm tr}(\omega^2)$ & $1$  & $\nu^0$\\
    $v^2\mathcal{S}$ & $1$  & $\nu^1$\\
    $v^2\mathcal{B}_\mu(\omega^2)^{\mu\nu}\mathcal{B}_\nu$ & $3$  & $\nu^3$\\  
    \hline
\end{tabular}  
\end{equation}

\subsection{Free-field checks}

For all free field theories in four dimensions, $\ln Z$ expanded at small $\beta$ truncates after ${\cal O}(\beta)$. (Equivalently, $S_q(x_1, x_2) = 0$ for $q \geq 3$.) Therefore, the only high-temperature relation to check is the one coming from $S_2(x_1, x_2)$ (i.e., from the ${\cal O}(\beta)$ piece in $\ln Z$) -- the other relations are all homogeneous and are therefore trivially satisfied for free theories. We compute these relations explicitly in the following subsections.

\subsubsection{Free conformal scalar}
The free scalar field theory has partition function:
\begin{align}
    \ln Z &= \frac{1}{(1-\omega_1^2)(1-\omega_2^2)} \times \Bigg[\frac{\pi^4}{45\beta^3} - \frac{\pi^2}{36\beta} (\omega_1^2 +\omega_2^2) + f^s(\omega_1^2,\omega_2^2) \nonumber \\ &~~~+ \left(\frac{1}{240} - \frac{1}{240} (\omega_1^2+\omega_2^2) - \frac{1}{240}(\omega_1^4 + \omega_2^4) - \frac{1}{144} \omega_1^2 \omega_2^2 \right) \beta + \text{non-pert}\Bigg] \nonumber \\
    &= -\frac1{\beta^3 x_1 x_2}\times \Bigg[-\frac{\pi^4}{45} + \frac{\pi^2}{36}(2-(x_1+x_2))\beta^2 - f^s(1-x_1,1-x_2)\beta^3 \nonumber \\ &~~~ + \left(\frac{7}{360} - \frac{7}{360}(x_1+x_2) + \frac1{240}(x_1^2+x_2^2)+\frac1{144}x_1x_2\right)\beta^4 + \text{non-pert}\Bigg]
\label{eq:scalar}
\end{align}
where $f^s(\omega_1^2, \omega_2^2)$ is some specific function that comes from the gapless sector; see \cite{Lei:2024oij} for its exact form.

From the ${\cal O}(\beta)$ piece of either line in (\ref{eq:scalar}) along with (\ref{relanom}) (or (\ref{explrel})), we get the $a$-anomaly:
\begin{equation}
a_{\text{scalar}} = \frac1{360}.
\end{equation}

\subsubsection{Free Dirac fermion}
For the free Dirac fermion theory on $S^3\times S^1$, the partition function is given by
\begin{align}
    \ln Z &= \frac{1}{(1-\omega_1^2)(1-\omega_2^2)} \times \Bigg[\frac{7\pi^4}{90\beta^3} + \frac{\pi^2}{36\beta} (-3+\omega_1^2 +\omega_2^2) + \left(\frac{17}{480} - \frac{1}{240} (\omega_1^2+\omega_2^2) - \frac{7}{480}(\omega_1^4 + \omega_2^4) + \frac{1}{144} \omega_1^2 \omega_2^2 \right) \beta + \text{non-pert}\Bigg] \nonumber \\
    &= -\frac{1}{\beta^3 x_1 x_2} \Bigg[-\frac{7\pi^4}{90} + \frac{\pi^2}{36}(1+x_1+x_2)\beta^2 +\left(-\frac7{1440}-\frac{19}{720}(x_1+x_2) + \frac7{480}(x_1^2+x_2^2) - \frac1{144}x_1x_2\right) \beta^4 + \text{non-pert} \Bigg].
    \label{eq:zdirac}
    \end{align}
From the $O(\beta)$ piece of either line in (\ref{eq:zdirac}) we then read off
\begin{equation}
a_{\text{Dirac fermion}} = \frac{11}{360}.
\end{equation}

\subsubsection{Maxwell theory}

For 4d free Maxwell theory, the partition function is:
\begin{align}
    \ln  Z &= \frac{1}{(1-\omega_1^2)(1-\omega_2^2)} \times \Bigg[\frac{2\pi^4}{45\beta^3} + \frac{\pi^2}{18\beta} (-6+5(\omega_1^2 +\omega_2^2))  -(1-\omega_1^2)(1-\omega_2^2)\ln \beta + f^v(\omega_1^2, \omega_2^2) \nonumber \\ &~~~~~~~+ \left(\frac{11}{120} - \frac{11}{120} (\omega_1^2+\omega_2^2) - \frac{1}{120}(\omega_1^4 + \omega_2^4) + \frac{11}{72} \omega_1^2 \omega_2^2 \right) \beta + \text{non-pert}\Bigg] \nonumber \\
    &= -\frac{1}{\beta^3 x_1 x_2}\Bigg[-\frac{2\pi^4}{45} +\frac{\pi^2}{18}\left(-4+5(x_1+x_2)\right)\beta^2 +(x_1 x_2 \ln \beta - f^v(1-x_1, 1-x_2))\beta^3 \nonumber \\ &~~~~~~~+\left(-\frac2{45} + \frac{2}{45}(x_1+x_2) + \frac1{120}(x_1^2+x_2^2) - \frac{11}{72}x_1x_2\right)\beta^4 + \text{non-pert}\Bigg]
    \label{eq:maxwellfree}
\end{align} 
where $f^v(\omega_1^2, \omega_2^2)$ is some specific function. Both it and the $\ln\beta$ term come from the gapless sector. Again from the $O(\beta)$ term we read off the $a$-anomaly:
\begin{equation}
a_{\text{Maxwell}} = \frac{31}{180}.
\end{equation} 

\subsubsection{$\mathcal{N}=4$ SYM at $\lambda=0$}

$\mathcal{N}=4$ SYM with gauge group $SU(N)$ at zero coupling is given by simply $6(N^2-1)$ copies of (\ref{eq:scalar}), $2(N^2-1)$ copies of (\ref{eq:zdirac}), and $N^2-1$ copies of (\ref{eq:maxwellfree}). This gives:
\begin{align}
    \ln Z &= \frac{N^2-1}{(1-\omega_1^2)(1-\omega_2^2)} \times \nonumber \\&\Bigg[\frac{\pi^4}{3\beta^3} + \frac{\pi^2}{6\beta} (-3+\omega_1^2 +\omega_2^2)  -(1-\omega_1^2)(1-\omega_2^2)\ln \beta + 6f^s(\omega_1^2, \omega_2^2) + f^v(\omega_1^2, \omega_2^2) \nonumber \\ &+ \left(\frac{3}{16} - \frac{1}{8} (\omega_1^2+\omega_2^2) - \frac{1}{16}(\omega_1^4 + \omega_2^4) + \frac{1}{8} \omega_1^2 \omega_2^2 \right) \beta + \text{non-pert}\Bigg] \nonumber \\
    &= -\frac{N^2-1}{\beta^3 x_1 x_2} \times \Bigg[-\frac{\pi^4}{3} +\frac{\pi^2}6(1+x_1+x_2)\beta^2 \nonumber + \big(x_1 x_2 
    \ln \beta -6f^s(1-x_1,1-x_2)-f^v(1-x_1,1-x_2)\big)\beta^3 \nonumber \\
    &~~~~~~~+\left(\frac1{16} - \frac18(x_1+x_2) + \frac1{16}(x_1^2+x_2^2) - \frac18x_1x_2\right)\beta^4 + \text{non-pert}\Bigg]
\end{align}
This gives
\begin{equation}
a_{\text{SYM}}  = \frac{N^2-1}{4}. 
\end{equation} 

\subsection{Six-dimensional free theories}

Free scalar theory:

\begin{align}
    &\ln Z(\beta;\omega_i) = \frac{1}{\prod_i(1-\omega_i^2)} \times \Bigg[\frac{2\pi^6}{945\beta^5} - \frac{\pi^4}{540\beta^3} \left(1+\sum_i\omega_i^2\right) + \frac{\pi^2}{2160\beta} \left(8 \sum_i \omega_i^2 + 3\sum_i \omega_i^4 + 5 \sum_{i < j} \omega_i^2 \omega_j^2\right) + f^s(\omega_1, \omega_2, \omega_3) \nonumber \\ &~~~~~~~~~~+ \frac{\beta}{60480}\Big(-31 + 31\sum_i \omega_i^2 + 52\sum_i \omega_i^4 + 77 \sum_{i< j} \omega_i^2 \omega_j^2 + 10 \sum_i \omega_i^6 + 21 \sum_{i\neq j} \omega_i^2 \omega_j^4 + 35 \omega_1^2\omega_2^2\omega_3^2\Big)\Bigg] + \text{nonpert} .
\label{eq:scalar6d}
\end{align}

\begin{align}
\frac{(-1)^{d/2}(d+2)}{2\Gamma(\frac{d+2}{2})}  \frac{1}{\prod_i(1-\omega_i^2)} &\frac{1}{60480}\Big(-31 + 31\sum_i \omega_i^2 + 52\sum_i \omega_i^4 + 77 \sum_{i< j} \omega_i^2 \omega_j^2 \nonumber \\&+ 10 \sum_i \omega_i^6 + 21 \sum_{i\neq j} \omega_i^2 \omega_j^4 + 35 \omega_1^2\omega_2^2\omega_3^2\Big)\Bigg|_{d=6, ~\omega_k^2 = e^{\frac{2\pi i k}{4}}} = \frac{1}{9072}~.
\end{align}

Free Dirac fermion:
\begin{align}
    &\ln Z(\beta;\omega_i) = \frac{1}{\prod_i(1-\omega_i^2)} \times \Bigg[\frac{31\pi^6}{1890\beta^5} - \frac{7\pi^4}{1080\beta^3} \left(5-\sum_i\omega_i^2\right)  + \frac{\pi^2}{4320\beta} \left(135 - 26\sum_i \omega_i^2 -21\sum_i \omega_i^4 + 10 \sum_{i < j} \omega_i^2 \omega_j^2\right) \nonumber \\ &+ \frac{\beta}{120960}\Big(-1835 -55\sum_i \omega_i^2 + 743\sum_i \omega_i^4 -14 \sum_{i< j} \omega_i^2 \omega_j^2 \nonumber + 155 \sum_i \omega_i^6 -147 \sum_{i\neq j} \omega_i^2 \omega_j^4 + 70 \omega_1^2\omega_2^2\omega_3^2\Big)\Bigg] + \text{nonpert} .
\label{eq:dirac6d}
\end{align}  
\begin{align}
\frac{(-1)^{d/2}(d+2)}{2\Gamma(\frac{d+2}{2})}  \frac{1}{\prod_i(1-\omega_i^2)} & \frac{1}{120960}\Big(-1835 -55\sum_i \omega_i^2 + 743\sum_i \omega_i^4 -14 \sum_{i< j} \omega_i^2 \omega_j^2 \nonumber \\ &+ 155 \sum_i \omega_i^6 -147 \sum_{ i\neq j} \omega_i^2 \omega_j^4 + 70 \omega_1^2\omega_2^2\omega_3^2\Big)\Bigg|_{d=6,~\omega_k^2=e^{\frac{2\pi i k}4}} = \frac{191}{45360}
\end{align}

\subsection{Holographic theories}

In $d$ dimensions, we get:
\begin{align}
\ln Z(\beta;\omega_k) &= \frac{\text{vol}~S^{d-1}(4\pi)^{d-1}\ell_{\text{AdS}}^{d-1}}{4d^d G_{d+1}} \frac{\beta^{1-d}}{\prod_i(1-\omega_i^2)}\Bigg[ 1 - \left(d^2(d-1) - d^2 \sum_i \omega_i^2\right) \left(\frac\beta{4\pi}\right)^{2} \nonumber \\
&+ \Bigg(\frac{d^3(d-1)(d-2)^2}{2} - d^3(d-2)^2\sum_i \omega_i^2 + d^3 (d-2)\sum_i \omega_i^4 + d^3(d-4) \sum_{i > j} \omega_i^2 \omega_j^2 \Bigg)\left(\frac\beta{4\pi}\right)^{4} \nonumber \\
&- \Bigg( \frac{d^4(d-1)(d-2)^3(d-5)}{6} - \frac{d^4 (d-2)^3 (d-5)}{2} \sum_i \omega_i^2 + d^4(d-2)(d^2-6d+10) \sum_i \omega_i^4 \nonumber \\
&~~~~+ d^4(d^3-11d^2+36d-40) \sum_{i>j} \omega_i^2 \omega_j^2 -d^4(d^2-12d+40)\sum_{i>j>k}\omega_i^2 \omega_j^2 \omega_k^2 - d^4(d^2-10d+20)\sum_{i\neq j}\omega_i^4 \omega_j^2 \nonumber \\
&~~~~-\frac{2d^4(d-2)(2d-5)}{3} \sum_{i}\omega_i^6
\Bigg)\left(\frac\beta{4\pi}\right)^{6} + \mathcal{O}(\beta^{8})
\Bigg].
\label{eq:ZAdSCFT}
\end{align}  

For $d=4$ and $d=6$, this matches the predictions. At $d=4$, we have 
\begin{align}
    \ln Z_{\rm Kerr-AdS_5} = &\frac{1}{G_5 (1-\omega_1^2)(1-\omega_2^2)} \times \Bigg[\frac{\pi^5}{8\beta^3} + \frac{\pi^3}{8\beta} (-3+\omega_1^2 +\omega_2^2)  + \left(\frac{3\pi}{16} - \frac{\pi}{8} (\omega_1^2+\omega_2^2) + \frac{\pi}{16}(\omega_1^4 + \omega_2^4) \right) \beta  \nonumber \\ &~~+
    \Bigg(\frac{1}{32\pi} - \frac{1}{32\pi} (\omega_1^2 +\omega_2^2) - \frac{1}{32\pi} (\omega_1^4 + \omega_2^4) + \frac{1}{32\pi} (\omega_1^6 + \omega_2^6) + \frac{1}{16\pi} \omega_1^2 \omega_2^2 - \frac{1}{32\pi} (\omega_1^2 \omega_2^4 + \omega_1^4\omega_2^2)\Bigg) \beta^3+ \mathcal O(\beta^5)\Bigg] \nonumber \\
    &= -\frac{1}{\beta^3 x_1 x_2 G_5} \times \Bigg[-\frac{\pi^5}{8} + \frac{\pi^3}{8}\left(1+x_1 + x_2\right)\beta^2 - \frac\pi{16}\left(1+x_1^2+x_2^2\right)\beta^4 \nonumber \\ &+\frac1{32\pi}\left(1-(x_1+x_2)-(x_1^2+x_2^2)+(x_1^3+x_2^3)+2x_1x_2-(x_1^2x_2+x_1x_2^2)\right)\beta^6 + \mathcal{O}(\beta^8)\Bigg].
\label{eq:holocft4}
\end{align}  
Each term in (\ref{eq:holocft4}) should be interpreted as having both quantum and stringy corrections. At leading order, we have $\frac{1}{G_5} = \frac{2N^2}{\pi}$. 

We see that the $\mathcal{O}(\beta^1)$ term in (\ref{eq:holocft4}) reproduces Table \ref{tab:examples} in the main text. The $\mathcal{O}(\beta^3)$ term is also instructive to check. We have
\begin{align}
    a_{0,0;3} &= \frac1{32\pi G_5} + \ldots, ~~~~~a_{1,0;2} = -\frac1{32\pi G_5}+\ldots, \nonumber \\
    a_{2,0;1} &= -\frac1{32\pi G_5}+\ldots, ~~~ a_{3,0;0} = \frac1{32\pi G_5} + \ldots, \nonumber \\
    a_{1,1;1} &= \frac{1}{16\pi G_5} + \ldots, ~~~~~ a_{1,2;0}=-\frac{1}{32\pi G_5} + \ldots,
    \label{eq:holo4dexphight}
\end{align}
which obeys
\begin{equation}
    6a_{0,0;3} + 12 a_{1,0;2} + 6 a_{2,0;1} + 6a_{1,1;1} = 0~.
    \label{eq:obeta3check}
\end{equation}
Just as in the $\mathcal{O}(\beta^1)$ piece, each term in (\ref{eq:holo4dexphight}) gets corrections from $\alpha'$ and $G_5$ (denoted with the $\ldots$), which must all cancel in (\ref{eq:obeta3check}). It would be interesting to compute these explicitly and verify the cancellation. 

In AdS$_7$, we get:
\begin{align}
    \ln Z(\beta;\omega_i) &= \frac{1}{G_{7}\prod_i(1-\omega_i^2)} \times \Bigg[\frac{4\pi^8}{729\beta^5} + \frac{\pi^6}{81\beta^3} (-5+\sum_i\omega_i^2) + \Big(\frac{5\pi^4}{27} - \frac{2\pi^4}{27} \sum_i\omega_i^2 + \frac{\pi^4}{54}\sum_i \omega_i^4 + \frac{\pi^4}{108} \sum_{i < j} \omega_i^2 \omega_j^2 \Big)\beta^{-1}  \nonumber \\ &+
    \Big(-\frac{5\pi^2}{54} + \frac{\pi^2}{18} \sum_i \omega_i^2 - \frac{5\pi^2}{72} \sum_i \omega_i^4 + \frac{7\pi^2}{216} \sum_i \omega_i^6 + \frac{\pi^2}{144} \sum_{i < j} \omega_i^2 \omega_j^2 - \sum_{i \neq j} \frac{\pi^2}{144} \omega_i^2\omega_j^4 + \frac{\pi^2}{144}\omega_1^2 \omega_2^2\omega_3^2\Big) \beta  + {\cal O}(\beta^3) \Bigg]
\label{eq:holocft6}
\end{align}  
and as a check
\begin{align}
\frac{(-1)^{d/2}(d+2)}{2\Gamma(\frac{d+2}{2})} & \frac{1}{G_{7}\prod_i(1-\omega_i^2)} \Big(-\frac{5\pi^2}{54} + \frac{\pi^2}{18} \sum_i \omega_i^2 - \frac{5\pi^2}{72} \sum_i \omega_i^4 + \frac{7\pi^2}{216} \sum_i \omega_i^6 \nonumber \\
    & ~~~+ \frac{\pi^2}{144} \sum_{i < j} \omega_i^2 \omega_j^2 - \sum_{i \neq j} \frac{\pi^2}{144} \omega_i^2\omega_j^4 + \frac{\pi^2}{144}\omega_1^2 \omega_2^2\omega_3^2\Big)\Bigg|_{d=6,~\omega_k^2=e^{\frac{2\pi i k}{4}}} = \frac{\pi^2}{48 G_{7}} = a_{6}. 
\end{align}

\subsection{Sketch of proof of the $a$-anomaly relation in arbitrary even dimensions}\label{anomproof}
Let us define for notational simplicity later on:
\begin{equation}
r:=e^{\pi i/(\kappa+1)},\qquad\zeta:=r^2=e^{2\pi i/(\kappa+1)}\,.
\label{eq:general-roots}
\end{equation}
We embed $S^{2\kappa-1}$ inside $\mathbb{C}^{\kappa}$ 
\begin{equation} z_a=\sqrt{\ell_a}\,e^{i\phi_a}~,
\qquad
\ell_a\ge0~,
\qquad
\sum_{a=1}^{\kappa}\ell_a=1~.
\end{equation}
We will call $\ell_a$s latitude co-ordinates while phases of $z_a$ form $T^{\kappa}$.

We write the metric in the following form:
\begin{equation}
 g=d \tau\otimes d\tau+
 \sum_{a=1}^{\kappa}
 \left(
 \frac{1}{4\ell_a}d\ell_a\otimes d\ell_a
 +\ell_ad\phi_a\otimes d\phi_a
 \right),
 \qquad
 \sum_ad\ell_a=0.
\label{eq:general-cylinder-new}
\end{equation}
At the cyclotomic point $\omega_j=r^j$, the thermal Killing vector is
\begin{equation}
 K=\partial_\tau-i\sum_{a=1}^{\kappa}r^a\partial_{\phi_a}~.
\label{eq:general-K-new}
\end{equation}
The physical thermal quotient is generated by $e^{\beta K}$. The large conformal diffeomorphism $\Sigma$ acting on the complexified cylinder $(\tau,\phi_1,\cdots,\phi_\kappa)$ is given by
\begin{equation}
		\begin{aligned}
			&\tau'=-i\phi_\kappa\,,\quad\phi_1'=-i\tau\,,\quad \phi_{a+1}'=\phi_a\,,\qquad
	 a=1,\ldots,\kappa-1\\
			&\ell_1'=\frac1{\ell_\kappa}\,,\quad 
			\ell_{a+1}'=-\frac{\ell_a}{\ell_\kappa},
			\qquad a=1,\ldots,\kappa-1.
	\end{aligned}
	\label{eq:general-Sigma-new}
\end{equation}
One can easily check that
\begin{equation}
	\Sigma^* g=-\frac{1}{\ell_{\kappa}}g\,,\qquad \Sigma_*K=r^{-1}K\,.
\end{equation}
If $K$ is restored to its canonical normalization then $\beta\mapsto r^{-1}\beta$.
Furthermore, we can easily show that the conformal diffeomorphism is isomorphic to $\mathbb{Z}_{2(\kappa+1)}$ since 
\begin{equation}
	\Sigma^{\kappa+1}=-\mathbf{1}\,.
\end{equation}

Now the idea is to represent the local hydrostatic contribution by an oriented $(2\kappa-1)$-form $\alpha_q[g,K]$ on any section transverse to the thermal flow.  We have
\begin{equation}
		\iota_K\alpha_q=0,
		\qquad
\mathcal{L}_K\alpha_q=0,
\label{eq:general-basic-new}
\end{equation}
The first of these implies that $\alpha$ does not have any component along $K^{\mu}$, while the second means that nothing changes as we move along the thermal orbit. Since the physical thermal data is $\beta K$,
\begin{equation}
		\alpha_q[g,\lambda K]=\lambda^{2q+1-2\kappa}\alpha_q[g,K].
\label{eq:general-K-hom-new}
\end{equation}
Using Weyl invariance of the EFT term, we can show that 
\begin{equation}
\Sigma^*\alpha_q= - r^{2q+1-2\kappa}\alpha_q
\label{eq:general-alpha-phase-new}
\end{equation}
where the $-$ sign comes from the orientation reversing nature of $\Sigma$.

Using the angular ($U(1)^{\kappa}$ acting on phases of $z_a$) invariance, we can construct a $(\kappa-1)$-form on the latitude space, defined by $\ell_a$s:
\begin{equation}
\omega_q:=\iota_{R_\kappa}\cdots\iota_{R_1}\alpha_q~,\qquad R_i:=\partial_{\phi_i}~.
\label{eq:general-omega-def-new}
\end{equation}
We can choose to perform the trivial angular integrations at this point and it does not affect the argument.

Because $\iota_K\alpha_q=0$, transformed angular vectors can be evaluated modulo $K$ and this induces a transformation on the angular variables:
\begin{equation}
B=
\begin{pmatrix}
0&0&\cdots&0&r\\
1&0&\cdots&0&r^2\\
0&1&\cdots&0&r^3\\
\vdots&&\ddots&\vdots&\vdots\\
0&0&\cdots&1&r^\kappa
\end{pmatrix},
\qquad
\det B=(-1)^{\kappa+1}r.
\label{eq:general-B-new}
\end{equation}
In other words, $B$ acts on the co-rotating angular coordinates.

Evaluating \eqref{eq:general-alpha-phase-new} on the $\kappa$ rotational vectors and $\kappa-1$ latitude tangent vectors gives
\begin{equation}
(\det B)\,\Sigma^*\omega_q=-r^{2q+1-2\kappa}\omega_q\,,
\end{equation}
which leads to 
\begin{equation}
\Sigma^*\omega_q=(-1)^\kappa\zeta^{q+1}\omega_q~.
\label{eq:general-omega-phase-new}
\end{equation}
For an oriented latitude $(\kappa-1)$-chain $\Delta:=\{\ell_a\geq 0\,, \sum_a \ell_a=1\}$, define
\begin{equation}
J_q[\Delta]:=\int_\Delta\omega_q~,
\end{equation}
which transforms as
\begin{equation}
J_q[\Sigma\Delta]
=(-1)^\kappa\zeta^{q+1}J_q[\Delta]~.
\label{eq:general-moving-chain-character-new}
\end{equation}
Recall that 
\begin{equation}
 x_a:=1-\zeta^a~,
 \qquad u:=\sum_{a=1}^{\kappa}x_a\ell_a=g(K,K)~.
\label{eq:general-u-new}
\end{equation}

All $x_a$ are nonzero. Hence we can introduce the projectively normalized coordinates
\begin{equation}
\chi_a:=\frac{x_a\ell_a}{u}~,
\qquad
\sum_{a=1}^{\kappa}\chi_a=1~.
\label{eq:general-chi-new}
\end{equation}
For real positive $x_a$ these coordinates map the physical latitude simplex to the standard simplex
\begin{equation}
\Delta_0=
\left\{
\chi_a\ge0,
\quad
\sum_a\chi_a=1
\right\}~.
\label{eq:general-standard-simplex-new}
\end{equation}
The large diffeomorphism induces the affine map
\begin{align}
\chi_1'&=x_\kappa\sum_{b=1}^{\kappa}\frac{\chi_b}{x_b}~,
\label{eq:general-chi-map1-new}\\
\chi_{a+1}'&=
\left(1-\frac{x_\kappa}{x_a}\right)\chi_a~,
\qquad a=1,\ldots,\kappa-1~.
\label{eq:general-chi-map2-new}
\end{align}
and the volume form 
\begin{equation}
\Omega_\chi:=d\chi_1\wedge\cdots\wedge d\chi_{\kappa-1}~,
\end{equation}
transforms as
\begin{align}
\Sigma^*\Omega_\chi=\zeta^{\kappa(\kappa-1)/2}\Omega_\chi=
(-1)^\kappa\zeta\,\Omega_\chi~.
\label{eq:general-Omega-phase-new}
\end{align}
Since $\omega_q$ is a top form on the latitude space, we can write 
\begin{equation}
\omega_q=\mathcal N_\kappa\,I_q(\chi)\,\Omega_\chi~,
\label{eq:general-omega-Iq-new}
\end{equation}
with $\mathcal N_\kappa$ independent of $\chi$. Hence we have 
\begin{equation}
I_q({\rm \Sigma}\chi)=\zeta^q I_q(\chi)~.
\label{eq:general-Iq-phase-new}
\end{equation}
Furthermore, note that 
\begin{equation}
    I_q(\chi) \in \mathbb C[\chi_1,\ldots,\chi_{\kappa-1}]~. 
\end{equation}
Now in what follows, we are going to show that 
\begin{equation}
	\int_{\Delta_0} \omega_q=0
\qquad\text{for}\qquad q\equiv\kappa\;(\mathrm{mod}\,\kappa+1)~.
\label{eq:general-vanishing-new}
\end{equation}
In order to do that, let us work in the affine hyperplane $\sum_a\chi_a=1$ and define the $\kappa+1$ complex walls
\begin{equation}
\mathcal W_0:=
\left\{
\sum_{a=1}^{\kappa}\frac{\chi_a}{x_a}=0
\right\}\,,\quad \mathcal W_a:=\{\chi_a=0\},
\qquad a=1,\ldots,\kappa~.
\label{eq:general-wall0-new}
\end{equation}
The affine map $\Sigma$ cyclically permutes these walls. For distinct labels $a,b\in\{0,1,\ldots,\kappa\}$ let $P_{ab}=P_{ba}$ be the intersection of all walls except $\mathcal W_a$ and $\mathcal W_b$.  Explicitly,
\begin{equation}
P_{0a}=e_a~,
\qquad a=1,\ldots,\kappa~,
\end{equation}
and for $a,b\ge1$,
\begin{equation}
P_{ab}:
\quad
\chi_a=\frac{x_a}{x_a-x_b}~,
\qquad
\chi_b=-\frac{x_b}{x_a-x_b}~,
\qquad
\chi_c=0\quad(c\ne a,b)~.
\label{eq:general-Pab-new}
\end{equation}
We orient 
\begin{equation}
\Delta_0=[P_{01},P_{02},\ldots,P_{0\kappa}]
\end{equation}
and $\Sigma$ acts as follows: 
\begin{equation}
\Delta_j:={\rm\Sigma}^j(\Delta_0)
=[P_{j,j+1},P_{j,j+2},\ldots,P_{j,j+\kappa}]~,
\label{eq:general-Deltaj-new}
\end{equation}
with labels understood modulo $\kappa+1$. 

\begin{remark}
As an example, for $\kappa=2$, we have
\begin{equation}
   \mathcal W_0=1+e^{2\pi i/3}\,,\quad \mathcal{W}_1=0\,,\quad \mathcal{W}_2=1 \,,
\end{equation}
Since there are three walls, intersection of all except two leaves one out, hence
\begin{equation}
    P_{01}=\mathcal W_2\,,\quad P_{02}=\mathcal W_1\,,\quad P_{12}=\mathcal W_0\,.
\end{equation}
Finally we have 
\begin{equation}
    \Delta_0=[1,0]\,,\quad\Delta_1=[1+e^{2\pi i/3},1]\,,\quad \Delta_2=[0,1+e^{2\pi i/3}]
\end{equation}
\end{remark}

Returning to the general proof, we note that for  any polynomial $F(\chi)$, we have 
\begin{equation}
    \sum_j (-1)^{\kappa j}\int_{\Delta_j} F(\chi)\Omega_\chi =0~.
    \label{id}
\end{equation}
For $\kappa=2$ this is exactly Cauchy's theorem around the triangle with vertices at $0,1$ and $1+e^{2\pi i/3}$.

Equation \eqref{eq:general-omega-phase-new} gives
\begin{equation}
J_q[\Delta_j]=
\left[(-1)^\kappa\zeta^{q+1}\right]^jJ_q[\Delta_0]~.
\label{eq:general-Jj-new}
\end{equation}
Applying \eqref{id}, we obtain 
\begin{equation}
0=\sum_{j=0}^{\kappa}(-1)^{\kappa j}J_q[\Delta_j]=\sum_{j=0}^{\kappa}\zeta^{j(q+1)}J_q[\Delta_0]~.
\label{eq:general-geometric-sum-new}
\end{equation}
If $q\equiv\kappa\pmod{\kappa+1}$, we have $(\kappa+1)J_q[\Delta_0]=0$ and hence \eqref{eq:general-vanishing-new} follows.

Finally, we recall that $S_q$ is proportional to $J_q$ except for $q=\kappa$ (we have not kept track of the factors arising due to trivial angular integrals above). For $q=\kappa$, we have an extra contribution coming from the Weyl non-invariant piece, proportional to $a$-anomaly, the proportionality constant can be fixed by comparing with the scalar field theory. 

\end{document}